\documentclass[letter]{ieice}
\usepackage[pdftex]{graphicx}

\usepackage[fleqn]{amsmath}
\usepackage{newtxtext}
\usepackage[varg]{newtxmath}
\usepackage{url}

\field{A}
\title{A Calculation Model for Estimating Effect of COVID-19 Contact-Confirming Application (COCOA) on Decreasing Infectors}
\authorlist{%
\authorentry{Yuto OMAE}{m}{af1}\MembershipNumber{1514610}
\authorentry{Jun TOYOTANI}{n}{af1}\MembershipNumber{}
\authorentry{Kazuyuki HARA}{m}{af1}\MembershipNumber{9999999}
\authorentry{Yasuhiro GON}{n}{af2}\MembershipNumber{}
\authorentry{Hirotaka TAKAHASHI}{m}{af3}\MembershipNumber{1721631}
}
\affiliate[af1]{The authors are with College of Industrial Technology, Nihon University, 1-2-1, Izumi, Narashino, Chiba 275-8575, Japan}
\affiliate[af2]{The author is with Nihon University School of Medicine, 30-1, Kami, Ooyaguchi, Itabashi, Tokyo 173-8610, Japan}
\affiliate[af3]{The author is with Research Center for Space Science, Advanced Research Laboratories, Tokyo City University, 8-15-1 Todoroki, Setagaya, Tokyo 158-8557, Japan}

\received{2020}{10}{x}
\revised{2020}{10}{x}

\begin{document}
\maketitle
\begin{summary}
As of 2020, COVID-19 is spreading in the world. In Japan, the Ministry of Health, Labor and Welfare developed COVID-19 Contact-Confirming Application (COCOA). The researches to examine the effect of COCOA are still not sufficient. We develop a mathematical model to examine the effect of COCOA and show examined result.
\end{summary}

\section{Introduction}
\begin{figure}[b]
  \begin{center}
    \includegraphics[scale=0.3]{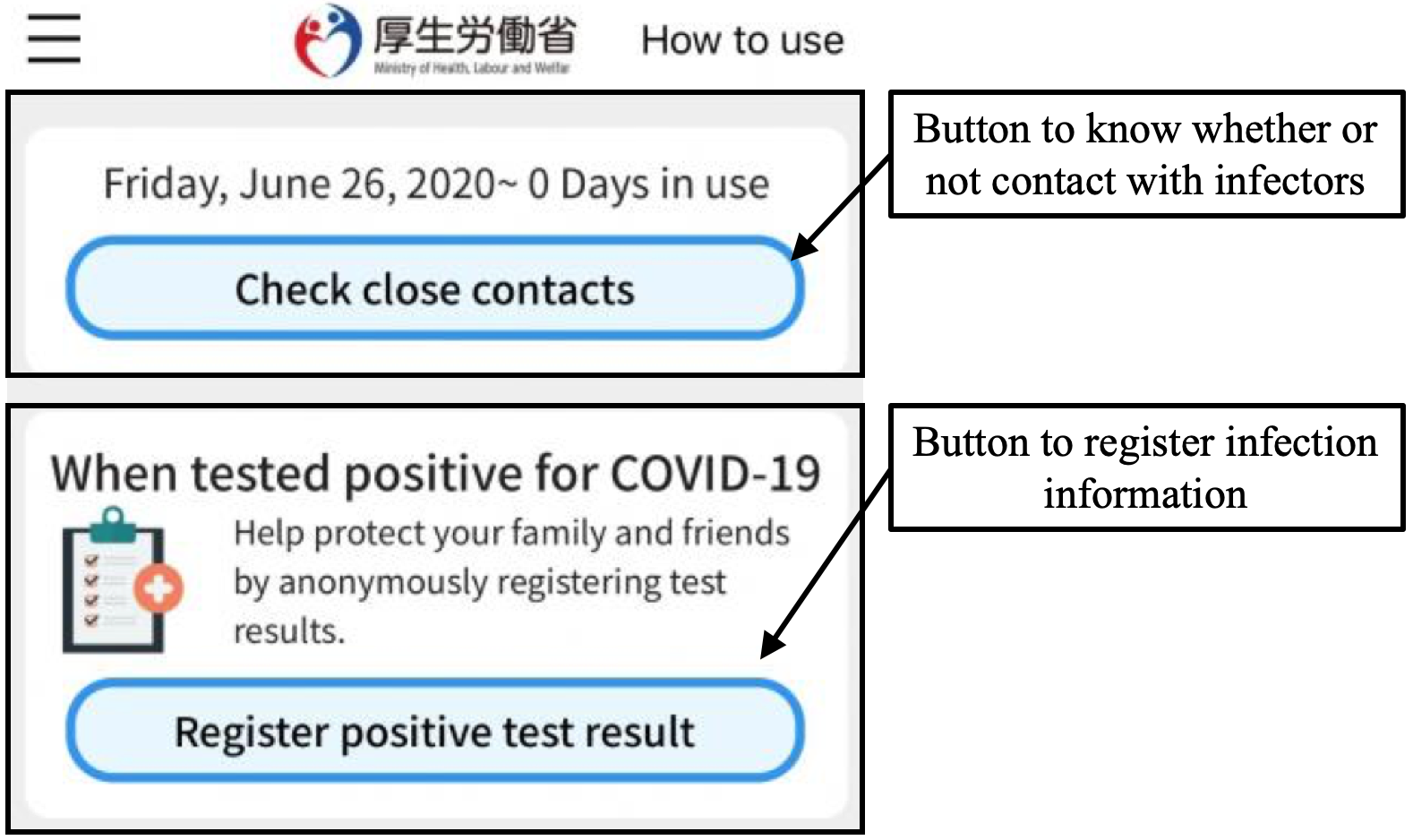}
    \caption{Schematic view of the COCOA~\cite{ref:usage}}
    \label{fig0}
  \end{center}
\end{figure}

On June 19, 2020, the Japanese government developed and released COVID-19 Contact-Confirming Application (COCOA)~\cite{ref:usage}, which was the smartphone app to decrease the number of COVID-19 infectors.
By using COCOA, the users can know whether or not they are contact with the infectors (refer to Fig.~\ref{fig0}).
If the close contact persons who receive the contact information from the app are staying at home, there is a possibility of decreasing the total infections (because they may be infections).

We consider that the reduction effect of new infectors increases as the usage rate of the app increases.
However, the usage rate of COCOA looks insufficient in Japan.
As of Oct. 2020, the number of install is about 18 million~\cite{ref:usage}.
In other words, the usage rate is about 15\%.
To increase the usage rate, it is important to report the reduction effect by the research (e.g. mathematical model and simulation etc.).

There are many researches on the model for estimating the number of COVID-19 infectors.
For example, Hou et al.~\cite{SEIR_C19_1} showed that a measure of decreasing the contact with the persons could effectively decrease the total infectors.
Chatterjee et al.~\cite{SEIR_C19_3} also conducted a simulation experiment as a case study in India.
The other simulation models of COVID-19 were also reported in~\cite{SEIR_C19_2}~\cite{SEIR_C19_4}~\cite{SEIR_C19_5}.
However, these researches did not show the reduction effect of the app such as COCOA.

The simulation model for estimating the effect of the app on decreasing infectors were developed.
For example, Hinch et al.~\cite{ref:hinch} used the individual-based network model and Omae et al.~\cite{ref:omae_arx} used the multi-agent simulation.
However, in the case of the infection disease estimation, the reliability verification of the simulation results is difficult because the researcher cannot experiment in the real world (i.e. we cannot calculate the differences between actual and estimated results).
As the alternative method to verify the reliability, it is necessary to examine the effects by the various simulation methods 
and verify that the obtained results are similar.
However, the research to survey the effect of the app such as COCOA to the number of total infectors is still insufficient.
Therefore, we develop a calculation model to know the app's effect in this letter.
We also examine whether or not the results are similar between previous researches and our model.

\section{Calculation model}
\subsection{Definition of the effect of COCOA}
We calculate the total infectors $I_t$ at the day $t$ as
\begin{eqnarray}
I_t &=& I_{t-1} + Rm^{-1}(1-p^2)I_{t-1} - m^{-1}I_{t-1} \nonumber \\
&=& I_{t-1}(1 + Rm^{-1}(1-p^2) - m^{-1}), \label{X_t}
\end{eqnarray}
where $R$ is a reproduction number. 
$R$ means the number of new infectors that the single infector infects other persons during total infection period.
$m$ is the number of average days for recovering from infection. 
Therefore, $m^{-1}$ in the third term is recovery rate of single day.
And $Rm^{-1}$ in the second term is the number of new infectors of single day that reproduced by one infector.
$p$ is the usage rate of the app in the total population.
The first term $I_{t-1}$ shows the total infectors at previous day.
The second term $Rm^{-1}(1-p^2)I_{t-1}$ shows the amount of new increased infectors.
COCOA can notice the contact information to a person who is contact with an infector, only if both use the app.
If only one uses the app, COCOA cannot notice the information to a close contact person.
Moreover, the app's usage probability of infectors is $p$.
And the app's usage probability of close contact persons is also $p$.
Because a joint probability of them is $p^2$, only this percentage of newly infectors receive contact information.
We assume that close contact persons who receive the contact information from the app do not go outside (i.e stay home).
In this case, they infect nobody.
Therefore, we can define the amount of new increased  infectors is $Rm^{-1}(1-p^2)I_{t-1}$.
The third term $m^{-1}I_{t-1}$ is the amount of decreased infectors by the recover or death.
Eq.(\ref{X_t}) was developed by partially referring to the equation for solving the number of infector in SIR model~\cite{ref:sir}.

In the case of using the number of initial infectors $I_0$, we can express $I_t$ from Eq.(\ref{X_t}) as:
\begin{eqnarray}
       I_t &=& I_{0}(1 + Rm^{-1}(1-p^2) - m^{-1})^t. \label{X_t2}
\end{eqnarray}
The proof of Eq.(\ref{X_t2}) is given in the appendix.

Then, the relative rate of the total number of infectors whether to use the app is $I_t(p)/I_t(p=0)$.
Therefore, the effect of the usage rate of the app $p$ on decreasing infectors can be defined as:
\begin{eqnarray}
E_t(p) &=& 1- \frac{I_t(p)}{I_t(p=0)} \nonumber  \\
&=& 1 - \frac{(1 + Rm^{-1}(1-{p}^2) - m^{-1})^t}{(1 + Rm^{-1} - m^{-1})^t},\label{E_t}
\end{eqnarray}

\subsection{Turning point by the usage rate of the app $p$ between infection spread and convergence}
In this section, we describe the turning point between infection spread and convergence by using the app.
Based on Eq.(\ref{X_t}), the coefficient of the amount increasing infectors is $Rm^{-1}(1-p^2)$, 
and the coefficient of the amount decreasing infectors is $m^{-1}$.
Therefore, if $Rm^{-1}(1-p^2) = m^{-1}$, the total infectors $I_t$ does not increase.
In other words, 
\begin{eqnarray}
Rm^{-1}(1-{p}^2) = m^{-1}, {\rm i.e.} \ R(1-{p}^2) = 1, 
\end{eqnarray}
is the condition of convergence of spreading infection disease.
We solve for $p$, then,
\begin{eqnarray}
{p} &=& (1 - R^{-1})^{1/2}. \label{eqc} 
\end{eqnarray}
Eq.(\ref{eqc}) means that, to lead the spreading infection to convergence, 
the required usage rate $p$ of app depends on the reproduction number of virus $R$.
The relationship between $p$ and $R$ by Eq.(\ref{eqc}) are shown in Table.\ref{tab1}.
As $R$ increases, the required $p$ also increases.

The reproduction number $R$ of COVID-19 is between 1.4 and 2.5~\cite{ref:rn}.
$R$ of SARS which is prevailed in 2003 in Hong-Kong is about 2.7~\cite{ref:sars}.
$R$ of seasonal influenza is about 1.3~\cite{ref:inf}.
Therefore, we consider that the app such as COCOA is effective to various infection disease.

Moreover, we consider the limitation of $p$ for $R$:
\begin{eqnarray}
\lim_{R \rightarrow \infty} {p} = \lim_{R \rightarrow \infty} (1 - R^{-1})^{1/2} = 1.
\end{eqnarray}
This means that even if the epidemic of infection disease of very high $R$ occuers, 
when everyone use the app i.e. $p=100\%$, the spread of infection will be overcame.
It is important to appropriately use the app for us.

Since our model is simple, there are some limitations.
For example, there is a time lag between an infector being infected and registering to the app in real world.
We did not consider it.
However, we emphasize that the model by simple calculation is important to understand the effect of COCOA 
on decreasing infectors

\begin{table}[tb]
\tabcolsep = 2.5pt
  \begin{center}
    \caption{Turning point between infection spread and convergence: the relationship between $R$ and $p$ in Eq.(\ref{eqc})}
    \begin{tabular}{c|cccccccc} \hline
$R$ & 1.0 & 1.25 & 1.50 & 1.75 & 2.00 & 2.25 & 2.50 & 2.75  \\ \hline
$p$ &  0.0\% &  44.7\% &  57.7\% &  65.5\% &  70.7\% &  74.5\% &  77.5\% &  79.8\%\\ \hline
    \end{tabular}
    \label{tab1}
  \end{center}
\end{table}

\section{Simulation}
\subsection{Condition}
To calculate the effect of the app by using our models, we set the following condition.
The initial infectors $I_0$ is 50 persons.
The maximum of simulation days is 50 days.
We consider the usage rates of the app: $p= 0, ~20, ~\cdots, ~80, ~100\%$.

We explain about $R$, which is the number of new infectors that the single infector infects other persons.
According to WHO, the reproduction number of COVID-19 is from 1.4 to 2.5~\cite{ref:rn}.
Therefore, we use $R=2.0$.

COCOA can send the notification of the contact information last 2 weeks~\cite{ref:usage}.
We assume the total infection period of COVID-19 is 2 weeks (14 days) i.e. $m=14$. 
In other words, we use the coefficient in the second term of Eq.(\ref{X_t}) $Rm^{-1}=2/14 \simeq 0.143$ and
the coefficient in third term of Eq.(\ref{X_t}) $m^{-1} = 1/14 \simeq 0.071$.


\subsection{Results and discussions}
\begin{figure}[t]
  \begin{center}
    \includegraphics[scale=0.6]{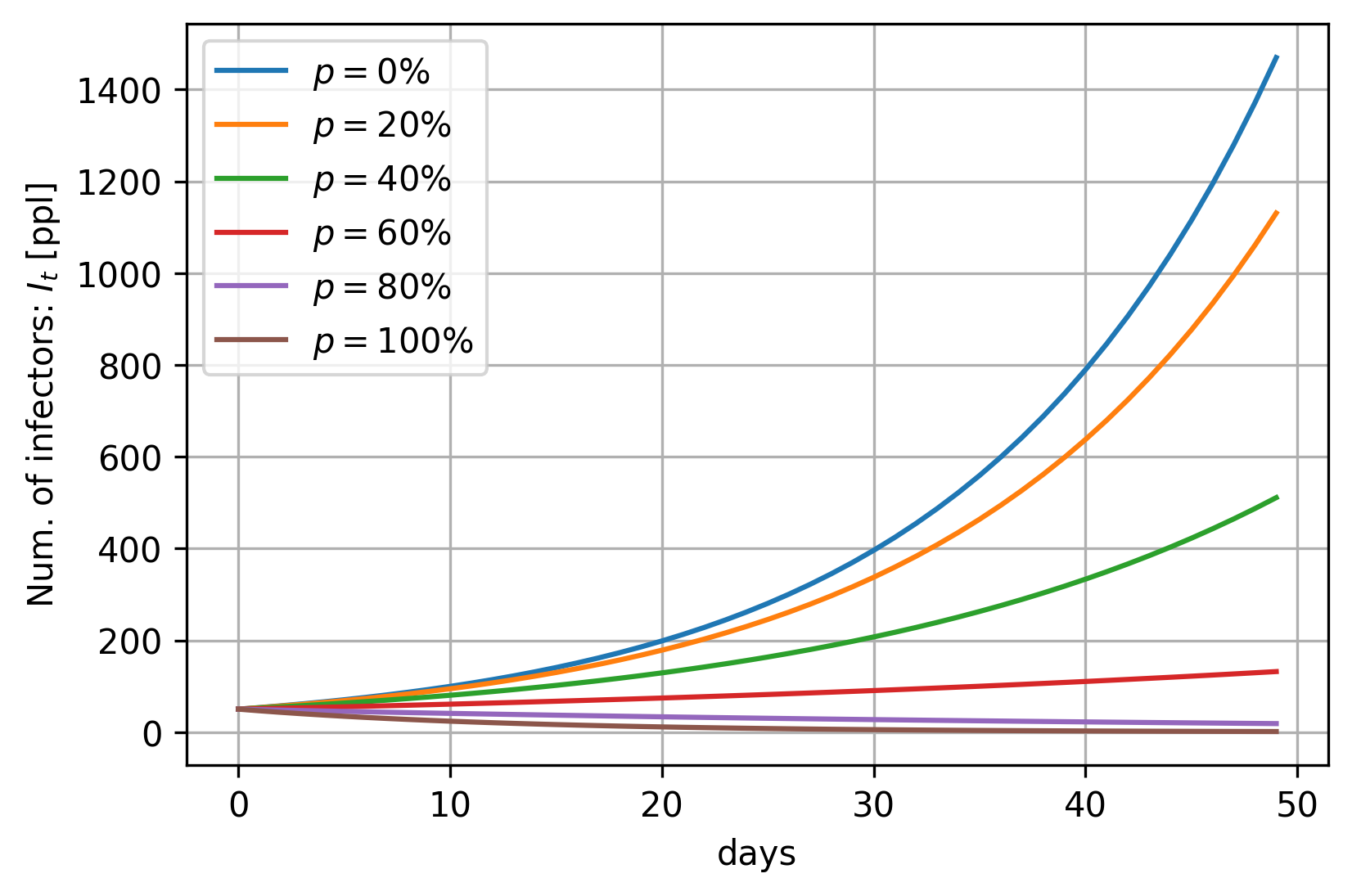}
    \caption{Total infectors $I_t$}
    \label{fig1}
  \end{center}
\end{figure}

The result of the calculation of the total infectors $I_t$ by Eq.(\ref{X_t2}) is shown in Fig.~\ref{fig1}. 
In the case of the usage rate $p=0\%$, the number of total infectors is over 1,400 persons.
Moreover, as $p$ increases, the total infectors decreases.
The most notable point is $p=60\%$.
In this case, the total infectors is nearly flat (slightly increased).
When $p=80$ and $100\%$, the total infectors decrease.
In other words, we can interpret that the spread of COVID-19 is the end.
This result has similar tendency to the result obtained by Hinch et al.~\cite{ref:hinch} based on the individual-based network model 
and Omae et al.~\cite{ref:omae_arx} based on multi-agent simulation.
They reported the spreading COVID-19 is convergence if the usage rate of the app over 60\%.
Therefore, we emphasize that our developed mathematical model supports the results of Hinch et al.~\cite{ref:hinch} and Omae et al.~\cite{ref:omae_arx}.
The point of appearing similar result by various methods is important.

\begin{figure}[t]
  \begin{center}
    \includegraphics[scale=0.6]{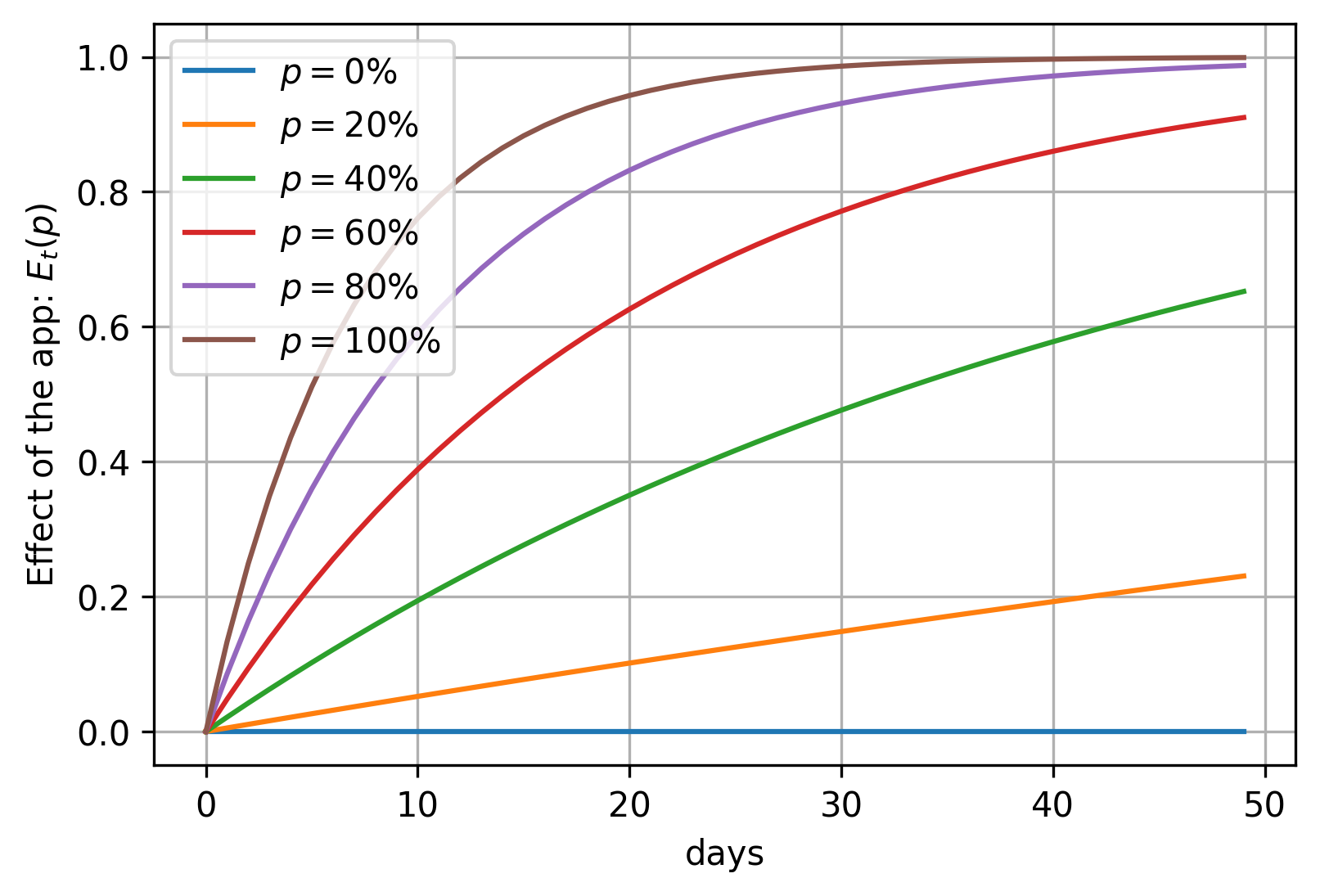}
    \caption{Effect of the app on reduction of infectors $E_t$}
    \label{fig2}
  \end{center}
\end{figure}

The effect of the usage rate of the app $p$ in $E_t$ (Eq.(\ref{E_t})) is shown in Fig.~\ref{fig2}.
As the basic trends, the reduction effect of the number of infectors increases over the time.
Moreover, as $p$ increases, the effect increases.
However, even if the usage rate of the app is low (e.g. $p=20\%$), the reduction effect appears steadily.

\section{Conclusion}
In this letter, we reported the effect of the app such as COCOA on decreasing infectors based on the simple calculation model.
As the result, we could understand the features/dynamics of the total infectors because we incorporated the usage rate of the app into the model.
However, other important parameters did not be incorporated and considered in this letter.
One of them is the registration rate of infection.
If infectors that use the app reject the registration of infection information, COCOA will not work.
Our model assume that all infectors who use the app register the infection information.
Moreover, we assume that close contact persons who received the contact information from the app do not go outside (i.e stay home).
We consider that some persons go outside without worrying about the contact notifications.
Thus, our developed model can be interpreted as the upper limit of the effect.
Therefore, we incorporate their points into the model as future works.
Finally, we will develop more desirable model to estimate the effect of COCOA.

\section*{Appendix: Proof of Eq.(\ref{X_t2})}
We explain the proof of Eq.(\ref{X_t2}) from Eq.(\ref{X_t}) of all natural number $t$ by using mathematical induction.

When $t=1$, Eq.(\ref{X_t}) is
\begin{eqnarray}
       I_1 &=& I_{0}(1 + Rm^{-1}(1-p^2) - m^{-1}), \label{1}
\end{eqnarray}
and Eq.(\ref{X_t2}) is
\begin{eqnarray}
       I_1 &=& I_{0}(1 + Rm^{-1}(1-p^2) - m^{-1}). \label{2}
\end{eqnarray}
Then, Eq.(\ref{1}) equals Eq.(\ref{2}) i.e. when $t=1$, Eq.(\ref{X_t}) equals Eq.(\ref{X_t2}).

After that, we assume Eq.(\ref{X_t}) equals Eq.(\ref{X_t2}) when $t=k$ ($k$ is a natural number).
In other words, 
\begin{eqnarray}
I_k &=& I_{0}(1 + Rm^{-1}(1-p^2) - m^{-1})^k, \label{3}
\end{eqnarray}
is true by Eq.(\ref{X_t2}).

When $t=k+1$, Eq.(\ref{X_t}) is
\begin{eqnarray}
           I_{k+1} &=& I_{k}(1 + Rm^{-1}(1-p^2) - m^{-1}) \nonumber \\
       &=& I_{0}(1 + Rm^{-1}(1-p^2) - m^{-1})^{k+1}. \label{4}
\end{eqnarray}
Note that we use Eq.(\ref{3}).

When $t=k+1$, Eq.(\ref{X_t2}) is
\begin{eqnarray}
       I_{k+1} &=& I_{0}(1 + Rm^{-1}(1-p^2) - m^{-1})^{k+1}. \label{5}
\end{eqnarray}
Thus, Eq.(\ref{4}) equals Eq.(\ref{5}) when $t=k+1$.

Therefore, Eq.(\ref{X_t}) equals Eq.(\ref{X_t2}) when all natural numbers $t$. 


\end{document}